\documentclass[twocolumn,superscriptaddress, longbibliography,
 amsmath,amssymb,
 aps, 
]{revtex4-2}
\usepackage{soul}
\usepackage{color}
\usepackage{graphicx}
\usepackage{dcolumn}
\usepackage{bm}
\usepackage{comment}
\usepackage{subfigure}
\usepackage{amsmath}
\usepackage[mathlines]{lineno}
\usepackage{nicefrac}

\newcommand{\aap}{Astron. Astrophys.}

\begin{document}

\preprint{APS/123-QED}

\title{Dynamo action driven by precessional  turbulence}

\author{Vivaswat Kumar}
 \email{v.kumar@hzdr.de}
 \affiliation{Institute of Fluid Dynamics, Helmholtz-Zentrum Dresden-Rossendorf, Bautzner Landstra{\ss}e 400, 01328 Dresden, Germany} 
 \affiliation{Institute of Process Engineering and Environmental Technology, Technische Universit\"at Dresden, 01062 Dresden, Germany}

\author{Federico Pizzi}
\affiliation{Institute of Fluid Dynamics, Helmholtz-Zentrum Dresden-Rossendorf, Bautzner Landstra{\ss}e 400, 01328 Dresden, Germany}
\affiliation{Department of Fluid Mechanics, Universitat Politècnica de Catalunya-BarcelonaTech,\\
Barcelona 08034, Spain}

 \author{George Mamatsashvili}
\affiliation{Institute of Fluid Dynamics, Helmholtz-Zentrum Dresden-Rossendorf, Bautzner Landstra{\ss}e 400, 01328 Dresden, Germany} 
\affiliation{E. Kharadze Georgian National Astrophysical Observatory, Abastumani 0301, Georgia}
 
\author{Andr\'e Giesecke}
\affiliation{Institute of Fluid Dynamics, Helmholtz-Zentrum Dresden-Rossendorf, Bautzner Landstra{\ss}e 400, 01328 Dresden, Germany}
 
\author{Frank Stefani}
\affiliation{Institute of Fluid Dynamics, Helmholtz-Zentrum Dresden-Rossendorf, Bautzner Landstra{\ss}e 400, 01328 Dresden, Germany}

\author{Adrian J.  Barker}
\affiliation{Department of Applied Mathematics, School of Mathematics, University of Leeds, Leeds, LS2 9JT, United Kingdom}

\date{\today}

\begin{abstract}
We reveal and analyze an efficient magnetic dynamo action due to precession-driven hydrodynamic turbulence in the local model of a precessional flow, focusing on the kinematic stage of this dynamo.  The growth rate of magnetic field monotonically increases with Poincar\'{e} number,  $\rm Po$, characterizing precession strength,  and magnetic Prandtl number, $\rm Pm$,  equal to the ratio of viscosity to resistivity, for the considered ranges of these parameters.  The critical ${\rm Po}_c$ for the dynamo onset decreases with increasing  $\rm Pm$. To understand the scale-by-scale evolution (growth) of the precession dynamo and its driving processes,  we perform spectral analysis by calculating the spectra of magnetic energy and of different terms in the induction equation in Fourier space. To this end, we decompose the velocity field of precession-driven turbulence into 2D vortical and 3D inertial wave modes. It is shown that the dynamo operates across a broad range of scales and exhibits a remarkable transition from a primarily vortex-driven regime at lower $\rm Po$ to a more complex regime at higher $\rm Po$ where it is driven jointly by vortices,  inertial waves and the shear of the background precessional flow. Vortices and shear drive the dynamo mostly at large scales, comparable to the flow system size, and at intermediate scales, while at smaller scales it is mainly driven by inertial waves. This study can be important not only for understanding the magnetic dynamo action in precession-driven flows,  but also in a general context of flows where vortices emerge and govern the flow dynamics and evolution.
\end{abstract}

\maketitle


\section{Introduction}\label{Intro}
Understanding the generation,  amplification and self-sustenance of magnetic fields in astrophysical and geophysical objects is the endeavour of dynamo theory \cite{roberts_1994,moffatt_dormy_2019, Rincon2019}.  From the magnetic field of our planet to distant stars and galaxies,  magnetohydrodynamic (MHD) dynamo models offer insights into the complex interplay between flows of conducting fluids and fields,  leading to the growth of the latter. To drive a dynamo, the kinetic energy of a flow must be transformed efficiently enough into magnetic energy, which in turn requires a strong driving mechanism for the flow.  

Among the known driving mechanisms for planetary dynamos, precession-powered motion is a complementary candidate \cite{Malkus1968, vanyo1991geodynamo} to the more generally accepted convection-driven ones \cite{Landeau2022}. Precession takes place when the rotation axis of a system periodically changes its orientation,  producing a body force that drives a flow in the (liquid) interior of the precessing body \cite{lagrange2011precessional}.  In particular, precession-driven flows are potentially able to convert large amounts of kinetic energy (up to $10^{11}-10^{21}$ W \cite{loper1975torque, rochester1975can, vanyo1991geodynamo}) to sustain the geomagnetic field \cite{Malkus1968}. This conversion can be due to instabilities \cite{lagrange2011precessional,giesecke2015triadic} that give rise to vortices and may also trigger turbulence.

Precession-driven flows have been studied both experimentally  \cite{noir2001experimental, Meunier2008, Goto2007, Herault2019, Kumar2023} and theoretically via numerical simulations in global settings \cite{kong2015transition, Marques2015, lin2016, Giesecke2018, Albrecht2018, Cebron2019, Lopez2019, Wu2020, Pizzi2021a, Pizzi2021b} to interpret those experiments. The capability of these flows to drive dynamos was demonstrated for laboratory flows in a specific parameter regime \cite{Tilgner2005, Wu2009, Cappanera2016, goepfert2016dynamos, Goepfert2019, Giesecke2018, Cebron2019, Kumar2023}.  An alternative approach widely used in studies of precessional flows in astrophysical and geophysical contexts  is a local model,  which describes the dynamics of a small segment of celestial bodies (stars,  gaseous planets or the liquid cores of rocky planets) in a rotating Cartesian coordinate frame \cite{Mason2002, Barker2016, Khlifi2018, pizzi2022interplay}.  In this model, the laminar background flow induced by precession was shown to be subject to a precessional instability \cite{Kerswell1993,Kerswell2002} and consequently this flow breaks down into a nonlinear (turbulent) state, composed of two basic modes -- two-dimensional (2D) vortices and three-dimensional (3D) inertial waves \cite{Barker2016,Khlifi2018,pizzi2022interplay}. As shown in these papers, the dynamical interplay among these two modes  and the background shear flow determines the sustenance and energetic balances of the precessional turbulence.

In the present paper, following our previous hydrodynamical study of the precessional flow dynamics in the local model \cite{pizzi2022interplay},  we consider the MHD extension of this model and investigate a magnetic dynamo action enabled by this flow,  focusing mainly on the kinematic stage of such dynamo. We demonstrate that this dynamo is powered by the precession-driven \textit{turbulence} resulting from the nonlinear saturation of the precessional instability and characterize the dependence of its growth rate on the main control parameters of the flow,  such as Poincar\'{e} number,  $\rm Po$,  which characterizes precession strength,  and magnetic Prandtl number, $\rm  Pm$,  equal to the ratio of viscosity to resistivity. To understand the scale-by-scale dynamics of the magnetic field, we perform spectral analysis, studying the evolution of the magnetic energy spectrum and mechanisms of its amplification in Fourier space as a function of these two parameters. Our most important finding is that there is a remarkable transition from a dynamo driven by 2D vortices at smaller $\rm Po$ to a one at higher $\rm Po$ driven jointly by 2D vortices,  background precessional flow shear and 3D inertial waves, which operate at different scales.

The paper is organized as follows: the physical model,  the main equations and the numerical setup are described in Section II. Analysis of the precession-driven dynamo in physical space is presented in Section III and in Fourier space in Section VI. Conclusions are given in Section V.

\section{Physical model and main equations}

In the local model, a precession-driven flow is considered in a Cartesian coordinate frame $(x,y,z)$ rotating around the vertical $z$-axis and precessing around another tilted $x$-axis with angular velocities $\Omega$ and ${\rm Po}\cdot\Omega$,  respectively,  where ${\rm Po}$ is the Poincar\'{e} number introduced above measuring precession strength.  In this frame,  precessional forcing gives rise to a laminar background flow with a linear shear along the $z$-axis, which oscillates in time and is proportional to $\rm Po$ \cite{Mason2002,Barker2016,pizzi2022interplay},
\[
\boldsymbol{U}_0= -2 \Omega\cdot {\rm Po}\cdot  z(\sin(\Omega t),\cos(\Omega t),0).
\]
The velocity perturbation $\boldsymbol{u}$ about this precession-driven base flow and the magnetic field  $\boldsymbol{B}$ are governed by the MHD equations for a conducting viscous and resistive incompressible fluid, which  in the rotating and precessing frame take the form \cite{Barker2016,pizzi2022interplay} 
\begin{multline}\label{eq:ns_pert}
(\partial_t +\boldsymbol{U}_0 \cdot \boldsymbol{\nabla}+\boldsymbol{u} \cdot \boldsymbol{\nabla})\boldsymbol{u} = - \frac{1}{\rho}\boldsymbol{\nabla}\Pi + \frac{1}{\mu_0\rho}(\boldsymbol{B}\cdot\boldsymbol{\nabla})\boldsymbol{B}+\nu \boldsymbol{\nabla}^{2} \boldsymbol{u} - \\ 2\Omega \boldsymbol{e}_z \times \boldsymbol{u} - 2\Omega \boldsymbol{\varepsilon}(t) \times \boldsymbol{u} +2\Omega u_z \boldsymbol{e}_z \times \boldsymbol{\varepsilon}(t),\end{multline}
\begin{multline}\label{eq:induction}
(\partial_t + \boldsymbol{U}_0 \cdot \boldsymbol{\nabla})\boldsymbol{B}= \nabla\times(\boldsymbol{u}\times \boldsymbol{B}) +\eta \boldsymbol{\nabla}^{2} \boldsymbol{B} -2\Omega B_z \boldsymbol{e}_z \times \boldsymbol{\varepsilon}(t), 
\end{multline}
\begin{equation}\label{eq:divergence}
\boldsymbol{\nabla} \cdot \boldsymbol{u}=\boldsymbol{\nabla} \cdot \boldsymbol{B}=0,
\end{equation}
where $\rho$ is the constant density of the fluid, $\Pi$ is the sum of thermal and magnetic pressures,  $\nu$ is the constant kinematic viscosity,  $\mu_0$ is the permeability of vacuum and $\eta$ is the constant magnetic diffusivity. The vector $\boldsymbol{\varepsilon}(t)={\rm Po}(\cos(\Omega t), -\sin(\Omega t),0)$ describes the effects of precession in these equations: Coriolis acceleration due to precession, $-2\Omega \boldsymbol{\varepsilon}(t) \times \boldsymbol{u}$,  and the stretching term $2\Omega u_z \boldsymbol{e}_z \times \boldsymbol{\varepsilon}(t)$ of the perturbation velocity due to the shear of the background flow $\boldsymbol{U}_0$ in Eq. (\ref{eq:ns_pert}), which jointly give rise to the precessional instability \cite{Kerswell1993,Kerswell2002,Sahli2010}.  A similar stretching term $-2\Omega B_z \boldsymbol{e}_z \times \boldsymbol{\varepsilon}(t)$ related to the shear in Eq. (\ref{eq:induction}) describes the growth of the magnetic field at the expense of free energy of the background flow.
\begin{figure}[t!]
  \centering
  \includegraphics[trim=0.5cm 0.5cm 0cm 0cm,scale=0.6]{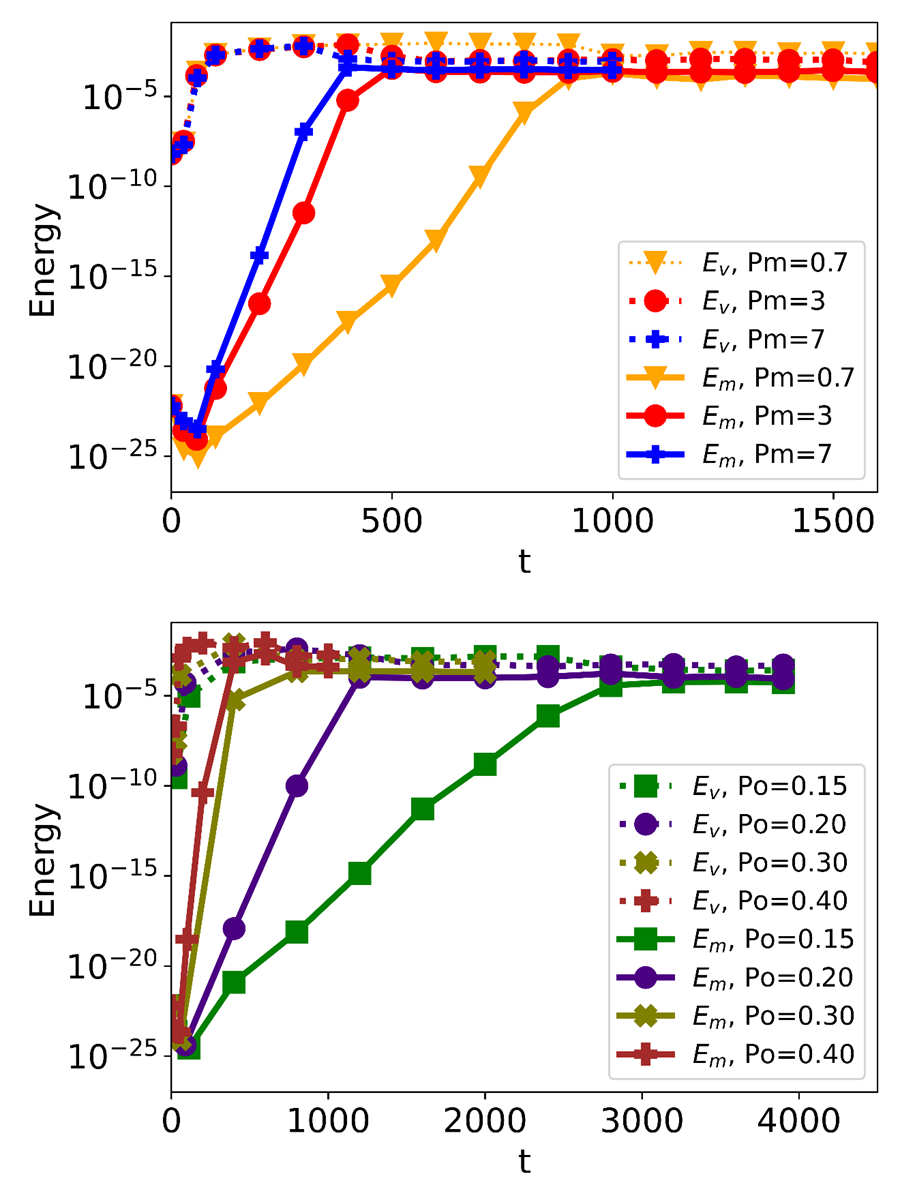}
  \caption{Evolution of the volume-averaged kinetic $\left< E_v \right>$ (dotted) and magnetic $\left< E_m \right>$ (solid) energies at $\rm Re=3\times 10^4$ for (a) different $\rm Pm$ at $\rm Po=0.3$ and (b) different $\rm Po$ at $\rm Pm=3$. The magnetic field energy starts to grow after the saturation of the precessional instability and increases exponentially by several orders of magnitude, indicating an efficient dynamo action due to precession-driven turbulence.} \label{fig:1}
\end{figure}

The flow is considered in a periodic cubic box with the same length $L$ in all directions,  $L_x=L_y=L_z=L$.  This box represents a small portion of a global precessional flow far from the actual boundaries of a system. In this case, we adopt mixed shear-periodic boundary conditions briefly summarized below, which are commonly used in the local model of the flow \cite{Barker2016,Khlifi2018,pizzi2022interplay}.
We cannot directly apply standard periodic boundary conditions, because the advection term $\boldsymbol{U}_0 \cdot \boldsymbol{\nabla}$ on the left hand sides of Eqs. (\ref{eq:ns_pert}) and (\ref{eq:induction}) depends linearly on $z$ due to the flow shear. To circumvent this, we transform $(x,y,z)$ coordinates to the  frame $(x', y', z')$ co-moving with the streamlines of the background flow $\boldsymbol{U}_0$, 
\begin{equation}\label{xcomoving}
x'=x-2{\rm Po}\cdot \cos(\Omega t) z,
\end{equation}
\begin{equation}\label{ycomoving}
y'=y+2{\rm Po}\cdot \sin(\Omega t) z,
\end{equation}
\begin{equation}\label{zcomoving}
z'=z.
\end{equation}
In this co-moving frame, the advection term is absent and we can impose usual periodic boundary conditions in the $(x',y', z')$-coordinates with the period lengths $L_x, L_y$ and $L_z$, respectively. After transforming back to the original  frame $(x,y,z)$, these boundary conditions take the periodic form in $x$ and $y$,
\begin{equation}\label{xboundary}
f(x,y,z,t)=f(x+L_x,y,z,t)~~~~(x~ {\rm boundary}),
\end{equation}
\begin{equation}\label{yboundary}
f(x,y,z,t)=f(x,y+L_y,z,t)~~~~(y~ {\rm boundary}),
\end{equation}
but shear-periodic form in $z$,
\begin{multline}\label{zboundary}
f(x,y,z,t)=\\f(x+2{\rm Po}\cdot \cos(\Omega t)L_z, y-2{\rm Po}\cdot \sin(\Omega t)L_z, z+L_z,t), \\(z~ {\rm boundary}),
\end{multline}
where $f\in \{\boldsymbol{u}, \Pi, \boldsymbol{B}\}$. These shear-periodic boundary conditions resemble those used in the widely known shearing box model of accretion disks \cite{Hawley_etal1995}. We note that the use of this local model with the associated shear-periodic boundary conditions is expected to be valid for modes with length-scales  smaller than the box size, but it becomes less suitable to describe larger structures comparable to the box size. Nevertheless, being devoid of real (rigid) boundary effects, this model is well suited for understanding the basic influence of precession on the dynamo action.

We normalize time by $\Omega^{-1}$, lengths by $L$, velocities by $\Omega L$, magnetic fields by $\Omega L (\mu_0\rho)^{1/2}$ and pressure by $\rho L^2\Omega^2$. With these normalizations, the kinetic and magnetic energy densities take the form $E_v= \boldsymbol{u}^2/2$ and $E_m=\boldsymbol{B}^2/2$, respectively. Besides the Poincaré number, the main parameters of the flow are the Reynolds number ${\rm Re}=\Omega L^{2}/\nu$ and magnetic Prandtl number  ${\rm Pm}=\nu/\eta$.  In real astrophysical and geophysical objects, precession is usually  weak and hence the Poincaré number is small $\rm Po \ll 1$,  while the Reynolds number is quite high \cite{Barker2016}.  In a first attempt to approach such a regime,  we fix ${\rm Re}=3\times 10^4$ and explore a broad range of ${\rm Po}=0.075-0.4$ and ${\rm Pm}= 0.1-10$, while the dependence on  $\rm Re$ will be analyzed elsewhere.

We solve Eqs. (\ref{eq:ns_pert})-(\ref{eq:divergence}) supplemented with the shear-periodic boundary conditions (\ref{xboundary})-(\ref{zboundary}) using the spectral code SNOOPY \cite{Lesur2005} adapted to the considered model of a precessional flow in Ref. \cite{Barker2016}.  Resolution for $\rm Pm < 1$, i.e., when the viscous scale is shortest,  is set as in \cite{pizzi2022interplay}: $(N_x,N_y,N_z)=(128, 128, 128)$ at $\rm Po \leq 0.125$ and $(256, 256, 256)$ at $\rm Po > 0.125$. On the other hand, for $\rm Pm > 1$, i.e., when resistive scale $\lambda_{\eta}$ is shortest instead, resolution is increased by a factor of ${\rm Pm}^{1/2}$.  This is because the corresponding resistive wavenumber $k_{\eta}=2\pi/ \lambda_{\eta}$ to be resolved in the code scales as $k_{\eta} \sim {\rm Pm}^{1/2}$ at $\rm Pm >1$ \cite{Rincon2019} and therefore resolution is adjusted accordingly. Solenoidal random velocity perturbations with rms $1.12\times 10^{-4}$ and a very small random (seed) magnetic field with rms $10^{-12}$ are imposed initially,  so that the back-reaction of the magnetic field on the flow remains negligible during the early growth phases of the field. This allows us to analyze the dynamics of the dynamo in the kinematic stage, which we mainly focus on in this paper. 

\begin{figure}[t!]
\centering
\includegraphics[trim=1cm 0.8cm 1cm 0cm,scale=0.5]{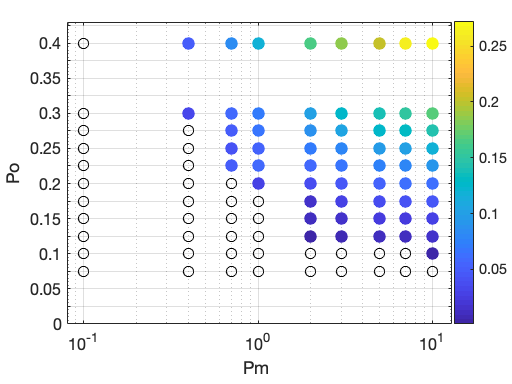}
\caption{Dynamo growth rate $\gamma$ in the ${\rm (Pm, Po)}-$plane at $\rm Re=3\times 10^4$ indicated by colored points,  whereas there is no dynamo at the empty points. The colored and empty points together represent all the simulation runs done in this work.}\label{fig:2}
\end{figure}

\section{Dynamo due to precession-driven turbulence}\label{Dynamo}

Figure \ref{fig:1} shows the evolution of the volume-averaged kinetic $\left< E_v \right>$ and magnetic $\left< E_m \right>$ energies at different $\rm Po$ and $\rm Pm$,  which indicates a two-staged dynamo process at work.  In the beginning, the kinetic energy grows exponentially as a result of the linear precessional instability,  while the magnetic energy decreases.  After several precession times $\sim (\epsilon\Omega)^{-1}$, the exponential growth saturates due to nonlinearity [advection term ($\boldsymbol{u} \cdot \boldsymbol{\nabla})\boldsymbol{u}$ in Eq.  (\ref{eq:ns_pert})] and,  as a result,  the flow settles down into a quasi-steady turbulence composed of 2D vortices and 3D inertial waves (\cite{Barker2016,pizzi2022interplay}, see Sec. \ref{spectral} below). Note that the dynamo action -- exponential growth of the magnetic field -- starts only \textit{after} saturation of the precession instability and is driven by the nonlinear (turbulent) velocity perturbations. After several hundred rotation periods, the magnetic field growth saturates nonlinearly due to the back-reaction of the Lorentz force on the flow,  which is discernible by the small drop in $\left< E_{v}\right>$ at the saturation point of $\left< E_{m}\right>$ (Fig. \ref{fig:1}).  The growth rate of the magnetic energy,  $\gamma=d~{\rm ln}\langle E_m\rangle/dt$,  in the kinematic regime and its nonlinear saturation level increase,  and hence the dynamo is more efficient, with increasing $\rm Po$ and/or $\rm Pm$. This dependence on $\rm Po$ and $\rm Pm$ is further explored in Fig.~\ref{fig:2}, which shows $\gamma$ in the $(\rm Po, Pm)$-plane and summarizes all of the simulation runs done. Note also in this diagram that the critical $\rm Po_c$ for the dynamo onset decreases with increasing $\rm Pm$. 
\begin{figure}[t!]
\centering
\includegraphics[trim=0.25cm 0cm 0cm 1cm,scale=0.53]{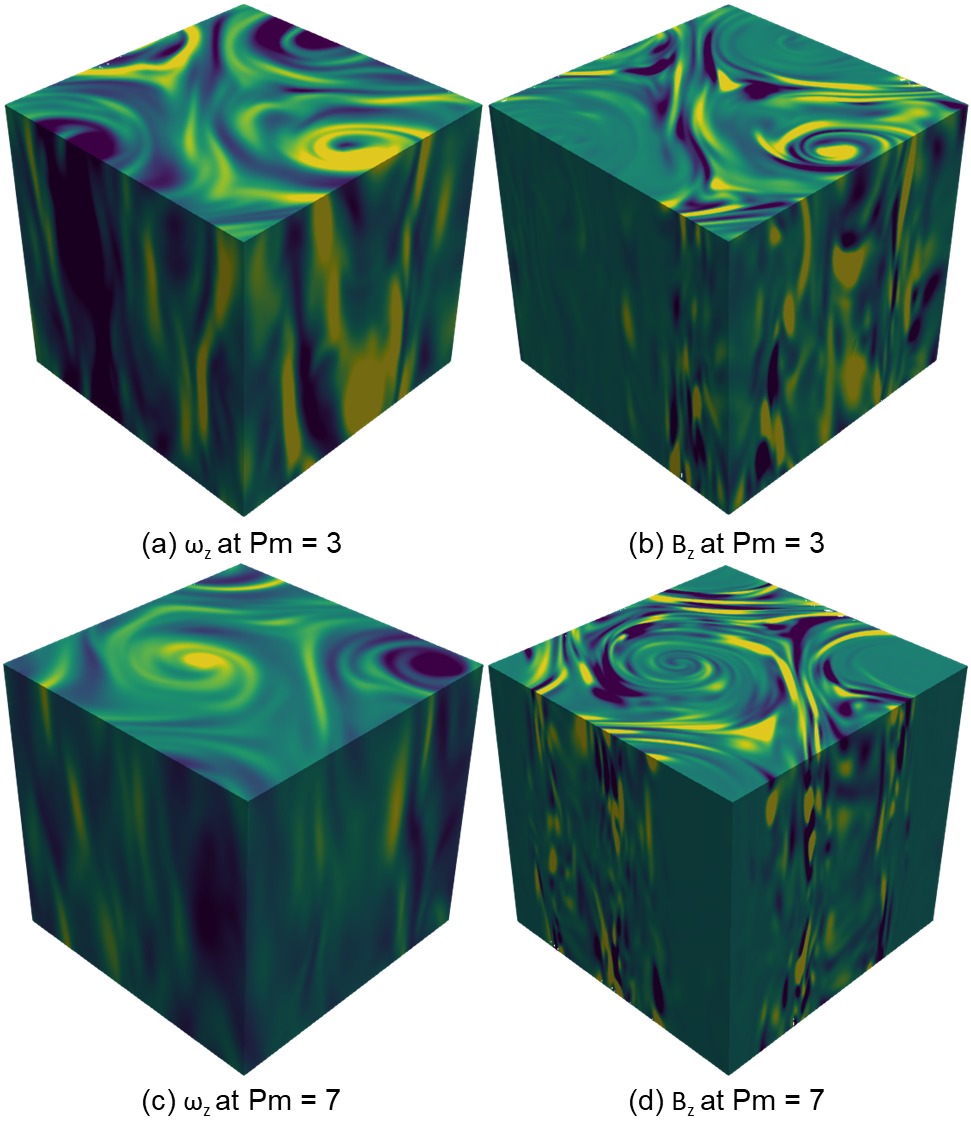}
\caption{Spatial distributions of the vertical vorticity $\omega_z=(\nabla \times \boldsymbol{u})_z$ at $\rm Po=0.15$ in two runs at (a) $\rm Pm=3$ and (c) $\rm Pm=7$ alongside the corresponding induced vertical field $B_z$ in (b) and (d), respectively.  For each $\rm Pm$,  the snapshots of $\omega_z$ and $B_z$ are shown at equal instants during the kinematic regime.  The characteristic length scale of the magnetic field structures is smaller than that of vortices and further decreases with increasing $\rm Pm$. }\label{fig:3}
\end{figure}

\begin{figure*}
\includegraphics[trim=10cm 5.8cm 10.5cm 1cm,scale=0.36]{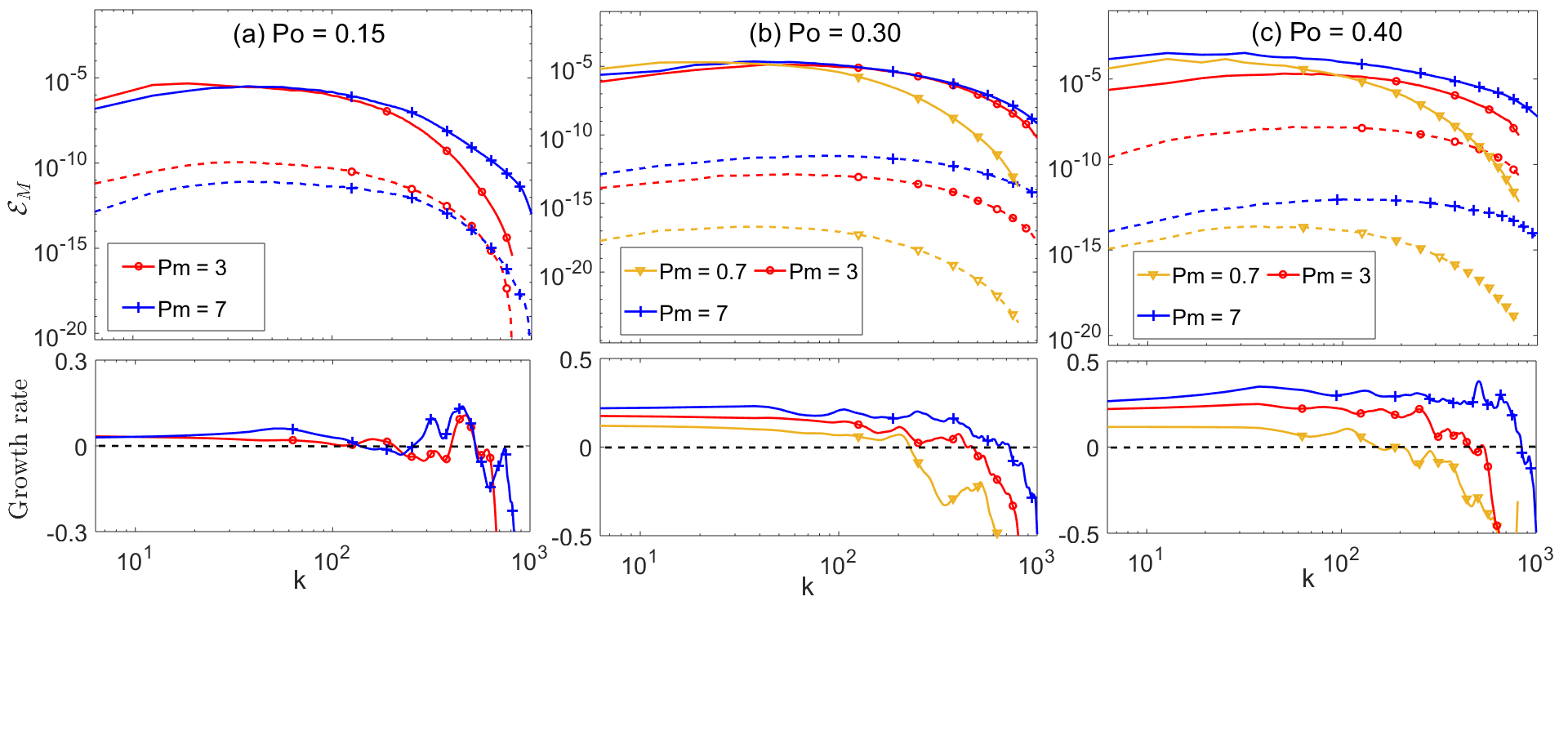}
\centering
\caption{(Top row) Shell-averaged magnetic energy spectra in the middle (dashed) and at the end (solid) of the kinematic stage of the dynamo at various $\rm Pm$ for (a) $\rm Po=0.15$,  (b) 0.3 and (c) 0.4. (Bottom row) The corresponding growth rate $\gamma(k)$ of the spectral magnetic energy vs. $k$.  There is no dynamo at $\rm Po=0.15$ and $\rm Pm=0.7$, hence the yellow curve is absent  in panel (a).}\label{fig:4}
\end{figure*}

Figure \ref{fig:3} shows the structures of the vertical vorticity $\omega_z=(\nabla\times\boldsymbol{u})_z$ in physical space at $\rm Po = 0.15$ and,  at the same instants, the induced vertical field $B_z$ during the kinematic stage in two runs at $\rm Pm= 3$ and $7$.  We observe vertically nearly uniform larger-scale columnar vortices embedded in a sea of smaller-scale 3D inertial waves,  as is typical of precession-driven turbulence \cite{Barker2016,Khlifi2018,pizzi2022interplay}.  The traces of these vortices are visible in the magnetic field,  since vortices are in fact the main drivers of the dynamo at this small $\rm Po$ (see below, Fig. \ref{fig:5}).  Note that although the magnetic field is induced mainly by these large-scale vortices,  the characteristic length scale of the field structures is still smaller than that of the vortices and decreases with increasing $\rm Pm$.  This behavior is also observed in the spectral analysis in the next section.

\section{Spectral analysis}\label{spectral}

To analyse dynamo action across length-scales,  we decompose the magnetic field into spatial Fourier modes both in the co-moving $(x',y',z')$ and original $(x,y,z)$ reference frames \cite{Barker2016,pizzi2022interplay},
\begin{multline}\label{eq:fourier}
\boldsymbol{B} = \sum_{\boldsymbol{k}} \bar{\boldsymbol{B}} \exp(ik_xx'+ik_yy'+ik_{z0}z') =\\
\sum_{\boldsymbol{k}}\bar{\boldsymbol{B}} \exp(ik_xx+ik_yy+ik_{z}(t)z).
\end{multline}
In the co-moving frame, the variables are all periodic and hence this decomposition has a standard form with the constant wavenumbers $k_x$, $k_y$ and $k_{z0}$, whereas in the $(x,y,z)$-coordinates, the vertical wavenumber $k_z$ varies with time. 
Indeed, using the invariance of the mode phases under the coordinate transformation, we can determine the wavenumbers in the original frame by substituting transformation (\ref{xcomoving})-(\ref{zcomoving}) in Fourier decomposition (\ref{eq:fourier}), regrouping the terms and equating the coefficients of $x,y$ and $z$ in the phases of both exponents.  As a result,  the horizontal wavenumbers $k_x$ and $k_y$, which do not depend on time, are the same in both frames, whereas the vertical wavenumber $k_z(t)$ oscillates in time about an average value $k_{z0}$,
\begin{equation}\label{vert_wavenumber}
k_z(t)=k_{z0}+2{\rm Po}(-k_{x}{\rm cos}(t)+k_{y}{\rm sin}(t)),
\end{equation}
thereby ensuring that the spatial Fourier modes satisfy the shearing-periodic boundary conditions (\ref{xboundary})-(\ref{zboundary}) in the $(x,y,z)$ frame ($\Omega=1$ in our units). Physically, the time-variation of $k_z$ is in fact brought about by the background shear flow $\boldsymbol{U}_0$, which advects the spatial Fourier modes (via the term $\boldsymbol{U}_0 \cdot \boldsymbol{\nabla}$), causing  the wavevector component along the direction ($z$) of the flow shear to change periodically in time.

As is commonly done in turbulence and dynamo theory,  below we will use the spherical shell-average of any spectral quantity $\bar{f}(\boldsymbol{k})$ (energy spectra, dynamical terms, etc.) in Fourier space, which at a time $t$ is defined in a standard way as $\sum_{k\leq |\boldsymbol{k}| \leq k+\Delta k}\bar{f}(\boldsymbol{k})$,  with the summation assumed over Fourier modes inside spherical shells with a given wavenumber magnitude (radius) $|\boldsymbol{k}|=k=(k_x^2+k_y^2+k_z^2)^{1/2}$ and width $\Delta k$ \cite{Alexakis2018}. Here $k_i=n_i\Delta k$,  $i\in \{x,y,z\}$, are the discrete wavenumbers in the cubic box with integer $n_i=0, \pm 1, \pm 2,..., \pm (N_i/2-1)$ and $\Delta k=2\pi/L$ is the grid cell size in Fourier space,  i.e.,  the minimum nonzero wavenumber in this box.

\begin{figure*}
\centering
\includegraphics[trim=5cm 1cm 16cm 0cm,scale=0.42,clip]{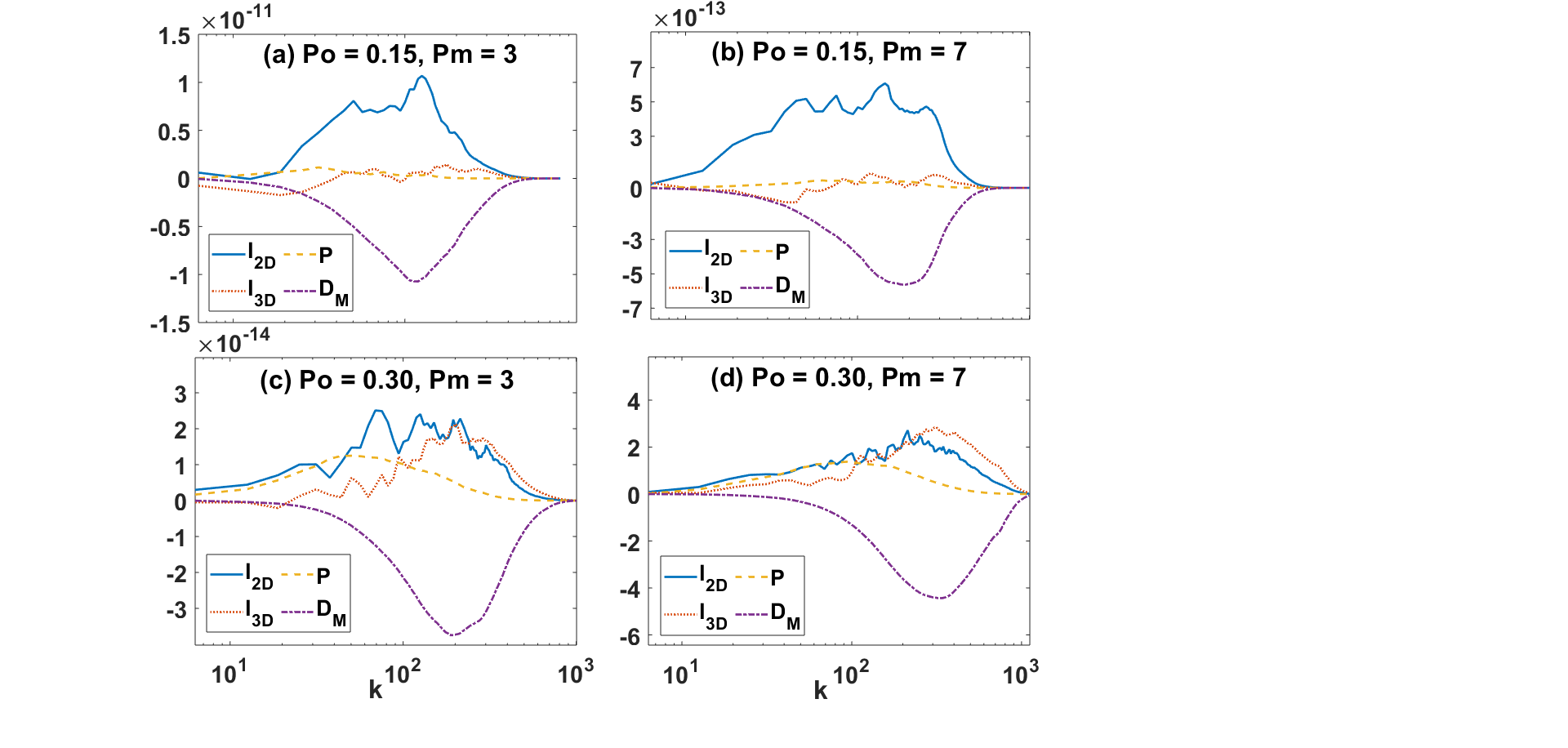}
\caption{Shell-averaged spectra of the dynamical terms in spectral magnetic energy Eq. (\ref{eq:magnetic_energy_fourier}): energy injection from the background shear flow, $P$, resistive dissipation, $D_M$, and the contributions from 2D vortices, $I_{2D}$, and 3D inertial waves, $I_{3D}$, in the middle of the kinematic stage at (a) Po = 0.15, Pm = 3; (b) Po = 0.15, Pm = 7; (c) Po = 0.3, Pm = 3 and (d) Po = 0.3, Pm = 7.}\label{fig:5}
\end{figure*}

Figure \ref{fig:4} shows the shell-averaged magnetic energy spectrum ${\cal E}_m(k)=|\bar{\boldsymbol{B}}|^2/2$  in the middle and at the end of the kinematic stage together with  the corresponding growth rate $\gamma(k)$ versus $k$ at different $\rm Po$ and $\rm Pm$. Note that in all the cases,  the growth rate is nearly constant and positive, $\gamma(k)>0$, at lower and intermediate $k\lesssim 100$, indicating the dynamo to be at work at these wavenumbers, and decreases, turning negative,  at higher $k$ due to resistive dissipation (see Fig. \ref{fig:5}). At small $\rm Po=0.15$, the growth rate weakly increases with $\rm Pm$,  mostly at higher wavenumbers [Fig. \ref{fig:4}(a)]. As a result,  the energy spectra at $\rm Pm=3$ and 7 have nearly the same shape and magnitude at small and intermediate wavenumbers, both during and at the end of the growth phase, while at high wavenumbers they are steeper  at $\rm Pm=3$ than at $\rm Pm=7$.  The magnetic energy spectrum and $\gamma(k)$ depend more strongly on $\rm Pm$ at higher $\rm Po=0.3$ and 0.4,  as seen in Figs. \ref{fig:4}(b) and \ref{fig:4}(c), respectively.  For a given $\rm Po$, the positive growth rate, increases with $\rm Pm$ at lower and intermediate $k$ and extends to higher $k$'s the larger $\rm Pm$ is, because of decreasing resistive dissipation with increasing $\rm Pm$ (at a given $\rm Re$). So,  the spectra grow faster and are shallower at high $k$ for higher $\rm Pm$.   

Comparing now the behavior of ${\cal E}_m(k)$  for different $\rm Po$ and a given $\rm Pm$ in Fig. \ref{fig:4},  we notice that with increasing $\rm Po$, $\gamma(k)$ moderately increases, although the maximum $k_m$ up to which there is still a dynamo does not appear to change much from $\rm Pm=0.7$ to 3. This critical wavenumber increases with $\rm Po$ at higher $\rm Pm=7$  because of stronger driving by waves at high $k$ (see below),  so the magnetic spectra are shallower at high $k$ for $\rm Po=0.3$ and 0.4 than those for $\rm Po=0.15$.                                       

To summarize,  in all the cases, the  magnetic energy spectrum spans a broad range of scales,  from the smallest wavenumbers corresponding to the system size up to the highest ones of resistive dissipation. Hence, the precession dynamo appears to be neither only large- nor only small-scale type.  The spectrum reaches a maximum at intermediate wavenumbers $10\lesssim k \lesssim 100$, with a specific value in each case depending on $\rm Po$ and $\rm Pm$ and being lower for lower values of these two parameters. Below we examine which modes of motion in the precession-driven turbulence  are mainly responsible for the dynamo growth at different $k$'s.

\subsection{Decomposition into vortical and wave modes}

The precession-driven hydrodynamic turbulence represents a special case of forced rotating turbulence, where precession acts as a forcing agent over a broad range of wavenumbers. Rotating turbulence has been very extensively studied in the literature (see a review \cite{Alexakis2018} and references therein). One of the main tools of analysis is the decomposition of a turbulent flow field into two basic types of perturbation modes -- vortices and inertial waves \cite{Smith1999, Buzzicotti2018,Alexakis2018}, which play key roles in the turbulence dynamics and energetic balances. An analogy between forced rotating turbulence and precession-driven turbulence, has motivated the application of the similar mode decomposition method to the latter \cite{Barker2016,Khlifi2018,pizzi2022interplay}. In the kinematic regime, Lorentz forces are negligible, so following those studies we can classify these two basic modes as done in the hydrodynamic case:
\begin{itemize}
\item
\textit{The vortical  modes} vary only in the horizontal $(x,y)$-plane and are uniform (aligned) along the vertical $z$-axis, having the form of a columnar vortex.  In the linear regime,  in the absence of precession,  these modes are strictly stationary with zero frequency,  $\omega=0$,  and in geostrophic balance.  However,  in the presence of precession,  they oscillate in time with a small amplitude around the stationary solution  due to the periodic variation of the background flow $\boldsymbol{U}_0$ (and hence $k_z(t)$).  The velocity of the vortical modes,  denoted as $\boldsymbol{u}_{2D}$,  typically has a dominant horizontal component over the vertical one, $u_h=(u_x^2+u_y^2)^{\nicefrac{1}{2}} \gg u_z$.  This \textit{slow} manifold is also referred to as 2D and three-component (2D3C) field in the literature,  since it varies only in $x$ and $y$ perpendicular to the rotation $z$-axis,  but still involves all three components of velocity.  Accordingly,  the Fourier transform of the velocity of the vortical mode,  $\bar{\boldsymbol{u}}_{2D}(k_x, k_y, t)$,  depends  only on $k_x$, $k_y$ and time $t$ with zero averaged vertical wavenumber,  $\langle k_z(t) \rangle=k_{z0}=0$.
\item
\textit{The inertial wave modes} vary both in the horizontal $(x,y)$-plane and along the $z$-axis. In the linear regime, they oscillate in time with nonzero frequencies $\omega=2\Omega k_z/k$ and are therefore considered to form the \textit{fast} manifold. In the presence of precession, inertial waves are driven by the precessional instability \cite{Kerswell1993,Kerswell2002} and grow, extracting energy from the background flow $\boldsymbol{U}_0$. The velocity of these modes, denoted as $\boldsymbol{u}_{3D}$, has in general comparable horizontal and vertical components, $u_h \sim u_z$, while its Fourier transform $\bar{\boldsymbol{u}}_{3D}$ has nonzero average $\langle k_z(t) \rangle=k_{z0}\neq 0$.
\end{itemize}

In the turbulent state, the dynamics of vortices and waves become coupled due to strong nonlinearity,  which ensures energy transfers between these two modes with different wavenumbers and hence variation of their amplitudes with time.  As a result,  the time-scales of the modes change,  forming approximately continuous spectra in frequency.  A detailed analysis of the energy spectra and  nonlinear dynamics of the vortical and inertial wave modes as well as energy transfers between them in physical and Fourier space in precession-driven turbulence was performed in \cite{Barker2016, pizzi2022interplay}. It was shown that for a given $\rm Re$,  at sufficiently low $\rm Po$, the dominant modes in the turbulent state are the vortical ones, as most of the turbulent energy resides in these modes. On the other hand, inertial waves become dynamically more important at higher $\rm Po$,  with their energy increasing relative to that of vortices, due to the stronger precessional instability driving these waves. We will show below that the vortical and wave modes play a central role in driving the precession dynamo and determining its spectral dynamics. In particular, the transition in the dynamo behavior with increasing $\rm Po$ is brought about by the changes in the hydrodynamical precession-driven turbulence regimes.

\subsection{Driving mechanisms of the precession dynamo}

To understand the roles of the background flow as well as the vortices and inertial wave modes in driving the precession dynamo, we first obtain the equation for the magnetic energy spectrum ${\cal E}_M$ by substituting Fourier transform (\ref{eq:fourier}) into Eq. (\ref{eq:induction}) and multiplying both sides by the complex conjugate $\bar{\boldsymbol{B}}^{\ast}$.  Then, in the electromotive force (EMF) $\nabla\times(\boldsymbol{u}\times \boldsymbol{B})$, the velocity is divided into vortical $\boldsymbol{u}_{2D}$ and wave $\boldsymbol{u}_{3D}$ parts,  $\boldsymbol{u}=\boldsymbol{u}_{2D}+\boldsymbol{u}_{3D}$,  giving
\begin{equation}\label{eq:magnetic_energy_fourier}
\frac{d{\cal E}_M}{dt}=P + D_M + I_{2D}+I_{3D},
\end{equation}
where $P=\Omega(\bar{\boldsymbol{B}}^{\ast}\bar{B}_z + \bar{\boldsymbol{B}}\bar{B}_z^{\ast})\cdot (\boldsymbol{\varepsilon}(t)\times\boldsymbol{e}_z)$ describes energy exchange between the magnetic field and the background flow due to shear; when $P>0$ energy is injected from the flow into the field. The second term $D_M=-2k^2{\cal E}_M/(Re\cdot Pm)<0$ is always negative and describes resistive dissipation. The third $I_{2D}$ and fourth $I_{3D}$ terms describe magnetic energy production, respectively, by vortical and inertial wave modes and are given by $I_{2D}={\rm i}[\bar{\boldsymbol{B}}^{\ast}\cdot\boldsymbol{k}\times(\boldsymbol{u}_{2D}\times\boldsymbol{B})_{\boldsymbol{k}}  - \bar{\boldsymbol{B}}\cdot\boldsymbol{k}\times(\boldsymbol{u}_{2D}\times\boldsymbol{B})_{\boldsymbol{k}}^{\ast}]/2$ and $I_{3D}={\rm i}[\bar{\boldsymbol{B}}^{\ast}\cdot\boldsymbol{k}\times(\boldsymbol{u}_{3D}\times\boldsymbol{B})_{\boldsymbol{k}}  - \bar{\boldsymbol{B}}\cdot\boldsymbol{k}\times(\boldsymbol{u}_{3D}\times\boldsymbol{B})_{\boldsymbol{k}}^{\ast}]/2$, where the subscript ``$\boldsymbol{k}$'' denotes the Fourier transforms of the EMF contributions coming from the velocity fields $\boldsymbol{u}_{2D}$ and $\boldsymbol{u}_{3D}$ associated with these two subsets of modes.

Figure ~\ref{fig:5} shows the shell-averaged spectra of these four terms $P(k), ~D_M(k), ~I_{2D}(k), ~I_{3D}(k)$ in the middle of the kinematic stage.  Depending on $\rm Po$ and $\rm Pm$,  either only vortices or jointly vortices, waves and the background shear are responsible for the magnetic field amplification.  At smaller $\rm Po=0.15$,  $I_{2D}$ is positive and much larger than $I_{3D}$ and $P$, i.e., $I_{2D} \gg P, I_{3D}$, implying that vortices predominantly drive the dynamo, whereas the driving by shear and inertial waves are relatively small [Figs. \ref{fig:5}(a) and \ref{fig:5}(b)]. As seen from this figure, this process is most efficient at intermediate $10 \lesssim k \lesssim 200$, where $I_{2D}$ is appreciable.  As $\rm Pm$ increases, $I_{2D}$ extends a bit to higher $k$, although its dependence on $\rm Pm$ is weak at this $\rm Po$,  consistent with the behavior of the magnetic energy spectra in Fig. \ref{fig:4}(a). It follows from this trend also that the total (over all $k$) driving by vortices is larger than that of inertial waves, $\int I_{2D}dk/\int I_{3D}dk= 7.78$ and 10.45 at $Pm=3$ and 7, respectively.

The dynamics changes qualitatively at higher $\rm Po=0.3$ -- the contribution of inertial waves and the shear in driving the dynamo becomes more appreciable relative to vortices,  since the corresponding terms $I_{3D}$ and $P$ are now comparable to $I_{2D}$,  as seen in  Figs.  \ref{fig:5}(c) and \ref{fig:5}(d) for the same $\rm Pm$, but they operate at different wavenumbers.  Specifically,  the waves amplify the magnetic field at higher $k\gtrsim 100$,  while the shear operates at intermediate wavenumbers $10 \lesssim k \lesssim 200$.  With increasing $\rm Pm$,  $I_{3D}$ dominates more over $I_{2D}$ and $P$ at large $k$, with its peak also shifting to higher $k$. So, at larger $Pm$, the 3D waves become the main driver of the dynamo at small scales,  while the vortices and the shear dominate at lower and intermediate $k$ [Fig. \ref{fig:5}(d)]. Accordingly, in this regime, the total driving by vortices and waves are comparable, $\int I_{2D}dk/\int I_{3D}dk= 1.06$ and 0.73 at $Pm=3$ and 7, respectively, although the second value is smaller due to the increased role of waves with increasing $Pm$.

Thus, Fig. \ref{fig:5} clearly shows a remarkable transition in the precession dynamo dynamics from a predominantly vortex-driven regime at lower $\rm Po=0.15$ to a more complex regime driven jointly by vortices, waves and shear at higher $\rm Po=0.3$.  This is actually related to the transition in the precession-driven turbulence regimes, as noted in Sec. IV.A, from the vortex-dominated state at lower $\rm Po$ to the state at higher $\rm Po$ where vortices and waves coexist and nonlinearly transfer energies between each other and the background flow.

\section{Conclusions}

In this paper, we have revealed and analyzed magnetic dynamo action powered by precession-driven \textit{turbulence}, which is capable of exponentially amplifying the magnetic field.  We showed that during the kinematic stage, the growth rate of the dynamo increases with the Poincar\'{e} ($\rm Po$) and the magnetic Prandtl ($\rm Pm$) numbers for the considered ranges of these two parameters.  The critical $\rm Po_c$ for the dynamo onset decreases with increasing  $\rm Pm$,  that is,  the dynamo sets in at lower $\rm Po$ the higher $\rm Pm$ is. 

Although our model is local,  it captures two basic ingredients of precessional turbulence: 2D vortices and 3D inertial waves.  Thus, this model is able to cover a broader range of length-scales than global ones -- from the system size, through intermediate scales mainly occupied by vortices, down to the shortest dissipation scales mainly occupied by inertial waves in our model. Previous works on the precession dynamo in global settings \cite{Tilgner2005,Wu2009,Cappanera2016,lin2016, Giesecke2018,Cebron2019, kong2015transition, goepfert2016dynamos, Goepfert2019, Kumar2023} emphasized the significance of large-scale vortices as a primary driver for large-scale magnetic field amplification, suggesting that the nature of the dynamo is closely linked to this flow pattern. This work,  going beyond those findings, has demonstrated for the first time the strong influence of the precession on the dynamo properties over a broad range of scales,  thereby providing  deeper insights into the dynamo action (growth rate, energy spectra, driving mechanism) over these scales. To achieve this, we have performed spectral analysis of the dynamics in Fourier space, quantifying the growth rate of the magnetic energy spectrum and the contributions from vortices, base flow shear and waves in driving the dynamo as a function of wavenumber. The main result of such a spectral analysis is a notable transition in the precession dynamo dynamics from a predominantly vortex-driven regime at smaller $\rm Po \lesssim 0.15$  to a more complex regime at larger $\rm Po \gtrsim 0.15$,  where vortices,  inertial waves and background shear act in cooperation to amplify the magnetic field at different scales. Vortices and shear drive the dynamo mostly at large (system size) scales and especially at intermediate scales, being therefore less sensitive to $\rm Pm$, while waves operate mainly at smaller scales, hence they depend more strongly on $\rm Pm$.  This dynamo transition is closely related to the intrinsic changes in the precession-driven turbulence regimes from the vortex-dominated one at smaller $\rm Po$ to a one at higher $\rm Po$ where vortices and inertial waves coexist, exchanging energy due to nonlinear transfers \cite{Barker2016,pizzi2022interplay}.

The sequence of processes leading to the dynamo action in the present problem (a precessional flow subject to instability $\rightarrow$ turbulence $\rightarrow$ vortices $\rightarrow$ dynamo) resembles that taking place in rotating convection.  As shown in  \cite{Guervilly2015},  thermal convection in the presence of rotation leads to turbulence and formation of large-scale vortices, which in turn amplify magnetic field as a result of induction (stretching) by vortical structures.  Thus, these two different (convection and precession) processes share a common generic mechanism -- \textit{vortex-induced dynamo}, which we have analyzed in this paper. Thus,  this study can be important not only for understanding the magnetic dynamo in precession-driven flows,  but also in a general context of flow systems where vortices emerge and govern the flow dynamics and evolution.

Our results can offer only qualitative insights regarding precession dynamo experiments due to differences in the regimes investigated and setups employed. Notably, in the present study, the magnetic Reynolds number $\rm Rm=\rm Pm\cdot \rm Re$, which controls dynamo onset and dynamics in those experiments, takes values $\rm Rm=3\times 10^3-3\times 10^5$, corresponding to the considered range of $\rm Pm=0.1-10$ and $\rm Re=3\times 10^4$, that far exceed the current capabilities of experimental facilities (e.g., DRESDYN dynamo experiment \cite{Pizzi2022mhd, stefani2014precession}). Additionally, in experiments one has to deal with real boundaries of the flow domain, which are excluded in our periodic box. The presence of boundary layers, as highlighted by Gans \cite{Gans_1971}, can be advantageous in experiments by lowering the $\rm Rm$-threshold of the precession dynamo and thus making it easier to excite due to shear in these layers, which, stretching magnetic field lines, expedites the  amplification of the field. What we can take away from this study as a guiding principle for future dynamo experiments  with precessional driving is that the developed turbulence within the bulk flow itself is conducive to dynamo action. So, an ideal experimental device should reach a high enough precession parameter $\rm Po$ to sustain both the vortices and inertial waves contained in this turbulence, which, as we showed, are capable of efficiently amplifying the magnetic field over a broad range of scales.

Finally,  we note that in the present paper, our main goal has been to uncover the dynamo action due to precession-driven turbulence and find conditions under which it exists. However, the ranges of $\rm Po$, $\rm Pm$ and $\rm Re$ considered here differ from those in precessing stars and planets. Still, our results can contribute to a better understanding of natural dynamos due to precession. The first step towards extrapolating the present analysis to relevant parameter regimes would be to investigate the effect of increasing $Re$ (which is huge in real astrophysical and geophysical objects, e.g., larger than $10^{10}$ in stellar interiors \cite{Viallet2013}) on the dynamo threshold, growth rate and driving mechanisms in the $(\rm Po, Pm)$-plane. This will allow us to see whether the precession dynamo can extend to even smaller $\rm Po$ and $\rm Pm$ (and larger $Re$) typical of weakly precessing stars and planets than those considered in this work.
\\
\\
\begin{acknowledgments}
This project has received funding from the European Research Council (ERC) under the European Union’s Horizon 2020 research and innovation program (Grant Agreement No. 787544), Shota Rustaveli National Science Foundation of Georgia (SRNSFG) [grant number FR-23-1277] and EPSRC grant EP/R014604/1. AJB was funded by STFC grants ST/S000275/1 and ST/W000873/1. He also thanks the Isaac Newton Institute for Mathematical Sciences, Cambridge, for support and hospitality during the programme ``Anti-diffusive dynamics: from sub-cellular to astrophysical scales'', where part of the work in this paper was undertaken. We thank anonymous Referees for useful comments and suggestions that improved the presentation of our results.
\end{acknowledgments}

\bibliography{biblio}
\end{document}